# Understanding the dynamics of biological colloids to elucidate cataract formation towards the development of methodology for its early diagnosis

SPYROS N. YANNOPOULOS[*] AND VASSILIKI PETTA

Foundation for Research and Technology Hellas – Institute of Chemical Engineering and High Temperature Chemical Processes (FORTH / ICE-HT), P.O. Box 1414, GR–26504, Patras, Greece

The eye lens is the most characteristic example of mammalian tissues exhibiting complex colloidal behaviour. In this paper we briefly describe how dynamics in colloidal suspensions can help addressing selected aspects of lens cataract which is ultimately related to the protein self-assembly under pathological conditions. Results from dynamic light scattering of eye lens homogenates over a wide protein concentration were analyzed and the various relaxation modes were identified in terms of collective and self-diffusion processes. Using this information as an input, the complex relaxation pattern of the intact lens nucleus was rationalized. The model of cold cataract – a phase separation effect of the lens cytoplasm with cooling – was used to simulate lens cataract at *in vitro* conditions in an effort to determine the parameters of the correlation functions that can be used as reliable indicators of the cataract onset. The applicability of dynamic light scattering as a non-invasive, early-diagnostic tool for ocular diseases is also demonstrated in the light of the findings of the present paper.

1. **Introduction and motivation in studying cataract formation**

The physics of colloidal suspensions has seen a great upsurge over the last two decades owing to the numerous theoretical and experimental investigations undertaken [1, 2]. The interest has arisen in view of the fact that colloidal suspensions serve as model systems in studies of changes in particles' interactions in dilute/dense systems, for example the sol-gel transition, as well as in studies of dynamically arrested matter, such as the liquid-glass-transition [3, 4]. Implicit in the vast majority of the studies of colloidal suspensions is the investigation of the effect of tuning interparticle interactions either by changing the colloidal particles concentration or by controlling certain properties of the solvent. The rich physical picture that has been gained by such studies has proved useful for understanding the behaviour of more complex systems, i.e. biological colloids. In the latter, and in particular in protein suspensions, one is frequently encountered with certain difficulties, i.e. departure from hard sphere behaviour, polydispersity, and multi-component mixture with highly asymmetric particle sizes, which are factors that complicate their study. This stands as an impediment in comprehending basic mechanisms of tissue functions and the routes to their degradation under pathological conditions. Such pathological conditions emerge frequently in cases where proteins aggregate or condense to form insoluble supra-molecular assemblies [5]. Even small amounts of aggregates can significantly alter the molecular structure which will modify protein function. Monitoring and understanding these effects are fundamental for resolving the relation of the molecular mechanisms that lead to proteins self-assembly and the effect that this could have to cellular function for the onset or the progress of a related disease.

---

[*] Corresponding author. E-mail: sny@iceht.forth.gr.

The ocular lens is presumably the most characteristic example of a mammalian tissue that exhibits complex colloidal behaviour. The lens can be considered as a dense aqueous suspension of globular proteins called crystallins (α-, β-, and γ-crystallins) whose total concentration ranges from ~200 mg ml$^{-1}$ in the lens periphery (cortex) up to ~500 mg ml$^{-1}$ in the lens nucleus. It is therefore quite surprising that such a concentrated suspension with a dense macromolecular content is highly transparent. Apparently, short-range interactions among crystallins play a decisive role in the underlying lens transparency [6, 7]. Lens opacification, usually referred to as *cataract*, reflects changes in interactions between crystallins and in particular is associated with changes in the short-range structural order. On general grounds, cataract is any opacification of the lens, which interferes with visual function by producing increased light scattering. Cataract is emerging as one of the most frequent diseases in humankind, being considered today as the most important cause of preventable blindness worldwide. Currently, cataract diagnosis is made clinically at the mature level of the disease. As the life expectancy is gradually increasing, age-related diseases like cataract will become more prominent among the population with high societal impact. It is therefore obvious the need for the development of a tool for early, non-invasive diagnosis of cataract onset.

2. **"*Pros* and *cons*" of dynamic light scattering in cataract studies**

A thorough understanding of the molecular mechanisms of lens opacification, signifying the onset of cataract, inevitably calls for an appreciation of the interactions between protein "particles" and the aggregation processes that take place upon aging and/or in the presence of other pathogenic factors. The key point to study interparticle interactions and aggregation is to investigate the *diffusion processes* of the colloidal particles at various time scales. Knowing that changes of density and/or concentration fluctuations underlie cataract formation and that the spatial dimensions of these fluctuations match to the visible light wavelength, it is rather apparent that dynamic light scattering (DLS) is an ideal tool for studying their dynamics. The added value of DLS studies of diffusion dynamics is that such studies can provide a simple, sensitive, and reliable methodology for non-invasive, early diagnosis of ocular diseases. The transparency of the lens and the other ocular tissues, situated externally to it, e.g. cornea, aqueous humour, as well as the dependence of the scattered intensity on the sixth power of the particle radius add to the advantages of using DLS as an *in vivo* diagnostic tool for cataract detection.

Research efforts in line with the aforementioned ideas were undertaken in the past mainly by Benedek and co-workers; see [7] for a review. However, several other authors have conducted similar light scattering and related spectroscopic studies as has been recently reviewed in [8] and [9]. In particular, DLS has been utilized for *in vivo* and *in vitro* studies of intact mammalian lenses. Despite the large body of experimental data the identification of the scattering elements in the lens and their role in the early stages development of cataract is still not clear. This is not unreasonable if we consider the very complex relaxation pattern which the lens nucleus exhibits even at physiological conditions as depicted in Fig. 1. Indeed, the high volume fraction of the various protein species that compose the ultradense, gel-like eye lens cytoplasm give rise to this broad autocorrelation function which can be described by at least four decay steps as shown by the distribution curve of relaxation times obtained by the inverse Laplace transform (ILT) method. The main difficulty is to reliably and uniquely rationalize these four modes in terms of the diffusive motions of the constituent protein "particles". The use of linear correlators of limited time scale span has prevented several researchers from obtaining the complete relaxation function over several time decades. This has stood as another obstacle for a correct analysis of those data which lack both the short-time plateau and the long-time baseline [10], see Fig. 1. Finally, based on the sizes and relative weight fractions of the various lens proteins (α-, β-, and γ-crystallins) it can be estimated that almost 85% of the scattered light intensity comes solely from the α-proteins; this is a further obstacle in DLS studies of eye lens proteins.

3.  **Identification of the lens scattering elements and the molecular mechanisms of cataract**

A cumbersome but efficient way to tackle the issue of identification of the diffusion processes of the complex relaxation function of the lens nucleus (shown in Fig. 1) is to follow a hierarchical approach where the lens proteins are isolated and studies of their dilute, semidilute, and dense suspensions are undertaken. Several DLS studies of lens protein suspensions' dynamics were carried out in the past and are still under consideration; these can be divided into two major classes. In the first one, proteins are separated into the various species and their suspensions are studied separately. In the second category, the total lens content (known as the lens homogenate) is diluted and hence the analogy of the three types of crystallins in the suspension is preserved similar to that of the intact lens. Although α-crystallins dominate in DLS, as mentioned above, and thus the dynamics of homogenate suspensions are dominated by the dynamics of this species, the study of homogenate suspensions is of paramount importance since *interactions between heterologous proteins have been considered as a decisive factor for the maintenance of lens transparency at high concentration* [6]. Despite this, studies of lens homogenates are very rare [8, 11[1]]. A detailed study of whole lens homogenates [8] enabled us to assign the four diffusive processes which correspond to the respective peaks in the ILT distribution shown in Fig. 1.

In brief, the fastest relaxation process originates from the *long-time, collective diffusion mode* of the proteins. This has been validated by experiments showing that the related diffusive motion becomes faster with increasing concentration [8]. From detailed analysis and using information from existing biochemical studies the second and third modes were identified with the *long-time, self-diffusion modes* of isolated α-crystallin particles and high-molecular weight assemblies of α-crystallins (HMα), respectively, present in the lens nucleus. Both modes exhibit strong volume fraction dependence, showing appreciable decrease with increasing volume fraction. The origin of the (fourth) slow mode is not obvious. This diffusion mode is reminiscent of the very slow structural relaxation modes observed in structural glasses termed long-range density fluctuations. It is thus reasonable to assume that in highly dense colloidal suspensions, approaching their glass transition, analogous modes may also exist. The volume fraction dependence of the diffusion coefficients was found to be satisfactorily described by recent predictions for colloids with soft character [12]. More details about the aforementioned can be found elsewhere [8].

The diffusion mode assignment presented above calls for a re-examination of the interpretation adopted frequently [7, 10] in efforts to identify a cataract index. In all previous studies fast and slowly diffusing modes were erroneously assigned to small and large scattering elements, respectively. This assignment has created confusion and has misled many authors to adopt the intensity ratio of these two modes as an indicator of cataract development. This assignment is incorrect for the following reasons. (a) The fast relaxation mode represents a collective diffusion process and hence it cannot be associated with an apparent particle size. (b) The slow mode cannot be associated with the large or HMα particles but rather with the native α-crystallins.

---

[1] In this work of lens homogenate studies [11] lens proteins were extracted from the periphery of the lens (the cortex), where γ-crystallins are largely absent. The presence of γ-crystallins is important in view of their specific role in cataract formation due to their attractive interactions.

## 4. Detection of the early stages of cataract in the intact lens nucleus

### 4.1. The "cold" cataract model

In order to study systematically the cataract onset in intact lenses using DLS one needs a way to induce and control the extent of cataract at *in vitro* conditions. An easily realizable model of cataract in the laboratory is the so-called cold cataract[2] [13, 14]. It can be easily induced by cooling the lens to temperatures below some characteristic or cold cataract temperature $T_{cc}$ located below the physiological one. $T_{cc}$ varies with the age and species of animal. Three main advantages of cold cataract over other types of cataract can be mentioned. (i) Cold cataract is induced gradually; this has the great advantage of offering a continuous monitoring of its stages due to the easy control of the external stimulus imposed, i.e., temperature. One is therefore able to examine the gradual appearance of the protein self-assembly process and thus to follow in detail the systematic changes of various spectral parameters of scattered light. (ii) In analogy to age-related cataract, cold cataract is of nuclear type. The fact that it does not form at the lens cortex facilitates laser light transmittance up to the lens nucleus. (iii) The effect is reversible; complete transparency can be achieved by warming the lens back to body temperature. Cold cataract is a physical aggregation driven mainly by hydrophobic interactions and is thus different from the senile cataract which is caused by new chemical covalent bonds between proteins. However, on a macroscopic scale, the formation of inhomogeneities and the onset of lens opaqueness which mind in DLS studies show some parallels in these two cases of cataract.

The cold cataract effect has its origin at the phase separation of the lens cytoplasm into a protein-poor and a protein-rich phase. The effect is analogous that that observed in water/polymer mixtures with a phase diagram exhibiting an upper critical solution temperature [15]. It is very interesting to remark that Bon [13] was the first to suggest that the reversible opacification of the lens during cooling could be related to a phase separation between two fluid phases, in analogy to what happens in binary mixtures that possess a critical mixing point. This suggestion has gone unnoticed is subsequent studies.

### 4.2. The methodology: Seeing the invisible

By employing the cold cataract model we have carried out a number of DLS studies of the phase separation effect in the nuclear part of intact porcine lenses *in vitro*. The study was conducted for several lenses in order to ensure the reproducibility of the results. Details about the experiment and the obtained results can found in [9]. We summarize below the main conclusions of this work. The data were evaluated with the aid of two analysis methods, i.e., using a sum of stretched exponential functions and the ILT technique, in order to extract with high accuracy the parameters related to the diffusion processes of protein particles and their aggregates. The first method employs the fitting of the electric-field time correlation function $g^{(1)}(q, t)$ with a discrete set of three stretched exponential (SE) functions, as described below:

$$g^{(1)}(t) = \sum_i A_i \exp[-(t/\tau_i)^{\beta_i}], \qquad i: f, sl, usl \qquad (1)$$

where $A_i$ is the amplitude or intensity of the *ith* decay step, $\beta_i$ is the corresponding stretching exponent, which is characteristic of the breadth of the distribution of the relaxation times and assumes values in the interval [0,1], and $\tau_i$ is the associated relaxation time. The indices "*f*", "*sl*", and

---

[2] The occurrence of the cold cataract phenomenon in mammal lens is known for more than 150 years; see the related references in [12].

"*usl*" denote the fast, slow, and ultraslow individual decay processes, respectively. It should be emphasized here that the limited time domain at which intensity autocorrelation functions were recorded in previous studies restricted the analysis to the use the inadequate scheme of two simple exponential terms, usually termed fast and slow [10]. The more detailed analysis [9] following the methods mentioned above revealed that the temperature dependence of many measurable parameters changes appreciably at the characteristic temperature ~16±1 °C for the porcine lens nucleus which is associated with the onset of cold cataract or phase separation of the lens cytoplasm.

Parameters that exhibit this change are the following: (i) The average scattered intensity at fixed temperature; (ii) the stretching exponent of the fast and slow components of the TCFs related to the heterogeneity of the local environments of the proteins and the polydispersity of the scatterers, respectively; (iii) the ratio of the scattering intensities associated with the fast and the slow components; (iv) the apparent activation energy of the fast and slow diffusion coefficients when plotted in an Arrhenius-type diagram; and (v) the apparent hydrodynamic radii of the scattering elements. While from the viewpoint of basic science all the above parameters give direct evidence for the onset of cold cataract, on practical grounds the use of many of them cannot lead to a viable solution to the problem of the early, noninvasive diagnosis of cataract at *in vivo* conditions. So far, cataract prediction has been based on the relative intensity ratio of the slow and fast processes, i.e., $A_{sl}/A_f$, which was erroneously considered as indicative of the population of the corresponding small and large particles. Our work showed that this ratio grows smoothly with decreasing temperature. Although this factor seemed a rather reasonable indicator of the formation of larger particles, in the spirit of the interpretation presented in [9] it is obvious that the $A_{sl}/A_f$ ratio cannot account for the aggregation process.

Compelling evidence for the inadequacy of the $A_{sl}/A_f$ ratio as a reliable indicator of cataract formation is provided by the data shown in Fig. 2. Time correlation functions were recorded at constant temperature – or equivalently at a certain extent of cataract – below $T_{cc}$ (at 10 °C) as a function of waiting time after equilibration at this temperature. Under these conditions the $A_{sl}/A_f$ ratio should remain constant over a broad range of $t^w$ times. However, as Fig 2(b) demonstrates, this ratio is appreciably scattered, whilst the stretching exponents which are related to the non-exponential character of the fast and slow process remain practically constant revealing the suitability of these parameters as indicators for monitoring the extent of cataract. Because the collective diffusion mode cannot be associated with single particle dynamics the increasing departure of $\beta_f$ from unity indicates an enhancement of the *spatial heterogeneity*, or alternative the generation of local microenvironments with distinct dynamical properties as it commonly occurs in supercooled liquids approaching their glass transition temperature. The development of spatial heterogeneity is indicative of the system's incipient phase separation with decreasing temperature that finally will engender lens opacification (cataract).

The analysis followed in [9] showed that the dynamics of proteins bears some resemblance to the dynamics of structural glasses where the apparent activation energy for particle diffusion increases below $T_{cc}$ indicating a highly cooperative motion. Application of ideas developed for studying the critical dynamics of binary protein/solvent mixtures, as well as the use of a modified Arrhenius equation, enabled us to estimate the spinodal temperature $T_{sp}$ of the lens nucleus.

## 5. Conclusions and Outlook

Studies on cataract issues have been focussed towards two main directions: (a) in the understanding the microscopic origin of cataract formation, and (b) in the development of an experimental methodology for early diagnosis of cataract *in vivo*. Eye lens protein suspensions have been frequently used as models of the lens and hence the physics of colloidal dynamics have proved valuable in appreciating selected aspects of cataract formation. The role of attractive interactions between heterologous eye lens proteins on lens transparency was emphasized recently [6]. In addition, the interest in studying the origin of cataract formation has recently revived in view of small-angle neutron scattering and simulation studies of eye lens protein mixtures [16][3] where it was shown that weak attractive forces between unlike proteins help to maintain lens transparency.

Light scattering-aided detection of cataract exhibits qualitative similarities and dissimilarities between cold cataract (physical, reversible aggregation and phase separation of lens proteins) and age-related cataract (chemical, irreversible binding of proteins through disulfide bonds). Although today we know to a good extent why and how some types of cataracts form [6, 7], the *in vivo* quantification of cataract index using DLS still remains a great challenge [17]. This has been the main impediment to the development of a reliable non-invasive optical diagnostic tool despite the many attempts for commercializing such a methodology (mainly USA patents). To overcome this obstacle we must determine those spectroscopic parameters that will be used as indicators for cataract; fundamental research is still needed towards this direction. In the present paper, we have briefly described work in our laboratory related to the aforementioned issues. DLS data of diffusive motions of lens homogenates, i.e. dilute and very dense suspensions of all kinds of proteins at relative concentrations found in the eye lens, were rationalized in analogy with dynamics of colloidal particles to identify the nature of diffusive motions. Then, based on these findings, DLS studies on several intact eye lenses led to the appreciation that the stretching (departure from exponential decay) of the two fastest decays steps in time correlation functions can indeed be used as sensitive and reliable indicators of qualitative and quantitative determination of cataract.

**Acknowledgements**

Financial support is gratefully acknowledged by the project "ΠΕΝΕΔ01/ΕΔ-559" which is co-funded by the European Union (European Social Fund) and the Greek State (Ministry of Development, GSRT) in the framework of the Operational Program "Competitiveness", Measure 8.3 – Community Support Framework 2000-2006.

---

[3] In this work mixtures of only α- and γ-crystallins were used. β-crystallins which amount to about 30 wt.% in the lens nucleus were missing; this is expected to cause significant changes in the balance of protein interactions in the mixture.

**Figure Captions**

Figure 1. A typical electric-field time correlation function of the porcine lens nucleus (open symbols) at the physiological temperature. The solid line is the best-fit curve obtained by either the ILT or SE analysis (see [8, 9] for details). The dashed-dotted line represents the ILT distribution of relaxation times. The solid circles curve is the corresponding experimental data of the bovine lens nucleus taken from [10], reproduced here for comparison.

Figure 2. (a) Dependence of the stretching exponents of the fast and slow processes of the porcine lens nucleus at 10 $^{\circ}$C as function of waiting time after equilibration at this temperature. (b) Corresponding dependence of the intensity ratio $A_{sl}/A_f$.

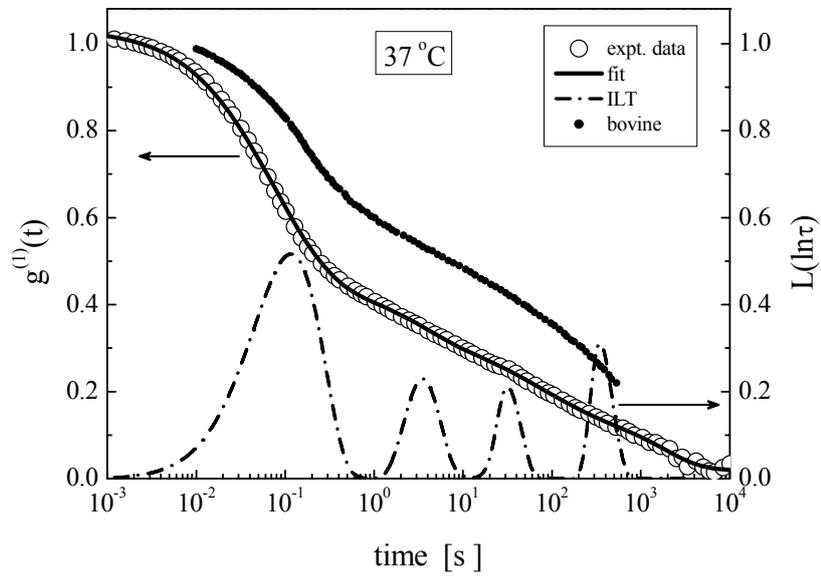

*Understanding dynamics of biological colloids…*: **Fig. 1**

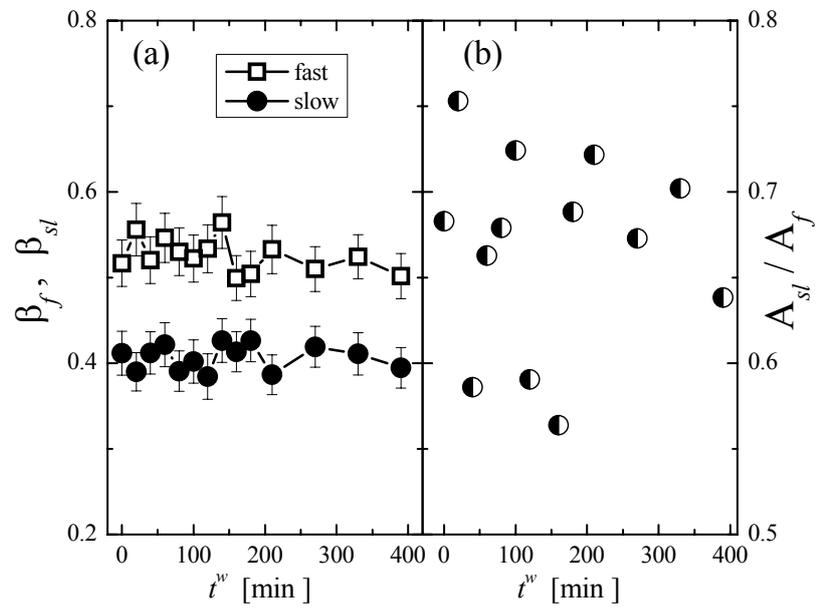

*Understanding dynamics of biological colloids…*:  **Fig. 2**